# Effect of crystallinity on the frictional and wear performance of molybdenum disulfide: A molecular dynamics study

Abhiram B R[a,*], Ilia Ponomarev[a], Tomas Polcar[a,b,*]

[a]*Dept. of Control Engineering, Czech Technical University in Prague, Prague 121 35, Czech Republic*
[b]*School of Engineering, University of Southampton, Highfield, SO17 1BJ, Southampton, UK*

**Abstract**

The frictional and wear performance of molybdenum disulfide ($MoS_2$) is significantly influenced by its intrinsic arrangement of crystals or crystallinity. In this study, we investigate the effect of crystallinty on coefficient of friction (COF) and wear in $MoS_2$ using a suite of reactive molecular dynamics (MD) simulations. A range of configurations, from amorphous to crystalline, is modeled to capture the effect of structural order on the tribological behavior. To study friction and wear, we simulate the sliding of a spherical rigid carbon body over the $MoS_2$ surface under varying crystallinity conditions. Our results reveal a pronounced reduction in COF with decreasing crystallinity, with crystalline $MoS_2$ exhibiting superlubricity. This behavior is attributed to the preservation of a flat sliding surface and frictional anisotropy, which enables lateral movement along low-resistance paths. In contrast, amorphous and polycrystalline $MoS_2$ with lower degrees of crystallinity displays a substantially higher COF, driven by increased surface roughness and atomic-scale energy dissipation. Furthermore, we examine the wear mechanisms under high normal loads, demonstrating that crystallinity enhances wear resistance by mitigating material deformation. These findings provide atomic-scale insights into the tribological performance of $MoS_2$, emphasizing the critical role of structural order in achieving ultralow friction. Our work corroborates with previous studies on superlubricity in $MoS_2$ and extends this understanding to rigid-body sliding conditions, offering valuable implications for designing low-friction and wear resistant solid lubricants.

## 1. Introduction

Transition metal dichalcogenides (TMDs) have shown exceptional mechanical [1], electronic [2], and tribological properties [3], making them suitable for a wide range of applications, including solid lubricants, energy storage devices, and nanoelectronics [4, 5]. Molybdenum disulfide ($MoS_2$), a twodimensional TMD, has attracted significant attention due to its unique combination of low friction [6, 7], high wear resistance [8], and thermal stability [9, 10]. These properties have led to the usage of $MoS_2$

*Corresponding authors*
Email addresses: binduabh@fel.cvut.cz (Abhiram B R), ponomili@fel.cvut.cz (Ilia Ponomarev), polcatom@fel.cvut.cz (Tomas Polcar)

in sectors such as aerospace and automotive industries [11, 12]. However, its performance significantly depends on crystallinity, which dictates the arrangement of atoms and layers within the material [13, 14].

Several experimental studies on $MoS_2$ have reported excellent mechanical properties, including elastic modulus [15], bending modulus [16] and fracture toughness [14]. The exceptional tribological performance of $MoS_2$ is attributed to its layered structure, where weak van der Waals interactions between layers enable easy shear [17]. This leads to low friction and superlubricity under specific conditions [6, 18]. Superlubricity, characterized by friction coefficients below 0.01 [19, 20], occurs when the sliding surfaces are atomically flat, leading to incommensurate contacts that minimize frictional forces [6]. However, the presence of humidity [21], oxygen [22], variations in external temperature [23], and degree of crystallinity [13, 14] govern the frictional and wear performance of $MoS_2$. Khare & Burris showed that the friction in $MoS_2$ increases with the presence of oxygen [24]. The negative influence of oxygen on the tribological properties has been attributed to the formation of molybdenum oxides on the sliding surface [24, 25]. Several experimental studies have also reported an increase in friction due to the presence of water [21, 26, 27]. John F. et al. studied the influence of microstructure on the oxidation and friction behavior of $MoS_2$ [28]. Recently, Liu et al. fabricated different crystal structures of $MoS_2$ by altering the grain boundaries [13]. Their study showed that a higher degree of crystallinity is beneficial for oxidation resistance and longevity of $MoS_2$. The wear study conducted by Hesam et al. on amorphous and polycrystalline structures revealed that the amorphous form exhibits wear resistance four times greater than that of its crystalline counterpart [29]. Although the effects of humidity and oxygen on the tribological properties have been extensively studied, the influence of varying degrees of crystallinity on its frictional and wear behavior remains poorly understood. A key challenge lies in the difficulty of fabricating $MoS_2$ with precise control over crystallinity [30, 31], which is essential for systematic investigation. Moreover, experimental techniques often face limitations in capturing the atomic-scale mechanisms governing friction and wear.

With the continuous advancement in computational power, researchers are increasingly utilizing computational techniques to study the mechanics of materials. Various computational methods, including molecular dynamics (MD) [32, 33], coarse-grained simulations [34], finite element methods (FEM) [35], and multiscale modeling, are employed to study material behavior across different length and time scales [36, 37], depending on the specific physics of interest. Among these approaches, MD simulations have proven to be particularly powerful in exploring atomic-scale phenomena, providing insights into fundamental mechanisms such as fracture, crack propagation, friction, and wear [38, 39]. Earlier MD simulations on $MoS_2$ have focused on estimating key mechanical properties including elastic modulus [40], ultimate strain [41], and fracture toughness [42]. For instance, Bao et al. investigated the crack propagation mechanism for single layer $MoS_2$ using atomistic simulations and proposed modifications to the Griffith criterion [43]. Additionally, MD simulations have been extensively used to study the influence of grain boundaries and defects on the mechanical properties and failure pattern in polycrystalline $MoS_2$ [31, 44, 45]. Recently, we employed MD simulations to study MoS$_2$ with varying degrees of crystallinity, demonstrating the dependence of fracture toughness and failure mechanisms on crystallinity [46].

In the context of tribology, significant progress has been made in understanding the friction and wear mechanisms of MoS$_2$ using MD simulations. For instance, Hu et al. studied the friction mechanism in multilayer MoS$_2$ under variable loads and shearing velocities [47]. Their results highlighted the irreversible deformation caused due to heavy load and shear velocity. Wei et al. explored the influence of grain boundary defects on the wear resistance through scratch simulations [48]. Other studies have examined the impact of interlayer spacing and temperature on frictional forces during sliding [49], as well as the effects of oxygen content in altering the tribological properties of MoS$_2$ [50]. Notably, Serpini et al. performed sliding simulation of ordered and disordered MoS$_2$, uncovering several nanoscale mechanisms for amorphous and crystalline configurations [51]. Additionally, studies have reported superlubricity in crystalline MoS$_2$ and detailed its nanoscale frictional behavior [52, 53]. Recently, research has focused on the abrasive wear due to sliding and rotating of a rigid body on crystalline MoS$_2$ [54]. Furthermore, previous studies have explored the influence of normal load and temperature on the wear mechanism during reciprocating friction [55]. However, despite these advancements, the influence of crystallinity on the frictional and wear properties of MoS$_2$ remains largely unexplored, highlighting a critical gap in the understanding of its tribological behavior.

In this study, we employ reactive MD simulations to systematically investigate the frictional and wear behavior of MoS$_2$ across a spectrum of crystalline configurations, ranging from fully amorphous to highly crystalline structures. Unlike previous studies, which have predominantly focused on a single crystalline form, our work uniquely explores the influence of varying degrees of crystallinity on tribological performance. We simulate sliding of a rigid body over MoS$_2$ surfaces under varying normal loads, enabling a comprehensive analysis of both friction and wear mechanisms. At low loads, we quantify the COF and elucidate the atomic-scale mechanisms governing frictional behavior. At higher loads, we examine wear pattern to understand material removal and deformation processes. By comparing our simulation results with experimental studies, we provide critical insights into the role of crystallinity in determining the tribological properties of MoS$_2$.

This paper is organized as follows: The methodology and MD simulation details are given in Section 2. The major outcomes from the study with a detailed analysis on COF, friction mechanism and wear characteristics are presented in Section 3. Finally, the concluding remarks from the study is provided in Section 4.

## 2. Modeling and simulation

In this section, we describe the modeling strategy used to create MoS$_2$ with varying degrees of crystallinity, along with the details of the MD simulations.

We modeled MoS$_2$ in various configurations, including crystalline, polycrystalline (PC), and amorphous structures. The PC models were created with 75%, 25%, and 10% crystallinity, denoted as PC75, PC25, and PC10, respectively. The initial configurations were constructed by randomly placing molybdenum (Mo) and sulfur (S) atoms within a simulation box of dimensions 63 × 40 × 35.5 Å$^3$. To achieve the desired polycrystalline structures, a sequence of annealing and quenching processes were employed, allowing control over the percentage of crystallinity. Detailed information regarding the annealing temperatures and the resulting crystallinity percentages are provided in Figure 1. For a better understanding, the atomic configuration of all PC structures are placed adjacent to the plots in the figure. The perfect crystal structure with six layers is created by orthogonalizing and replicating the unit cell in X and Y directions. Here, the stable 2H phase of MoS$_2$ is chosen, where each Mo atom is covalently bonded to six S atoms, and each S atom is covalently bonded to three Mo atom [56]. All models were designed to maintain a constant total number of atoms. The equilibrated structure of PC25 is shown in Figure 2a. The red and blue colored balls represent Mo and S atoms. Following equilibration, the simulation cell is extended in the Z direction to 75 Å to accommodate a rigid body. A spherical rigid body with a diameter of 20 Å, composed of 1431 atoms is positioned on top of the MoS$_2$ surface. The atomic configuration of the MoS$_2$ model with the rigid body is illustrated in Figure 2b. The density of crystalline and PC MoS$_2$ after equilibration is obtained as 5.0 and 4.8 g/cc, which is in accordance with the values observed in experimental studies [57]. The strategy adopted to measure degree of crystallinity is given in detail in Appendix A.

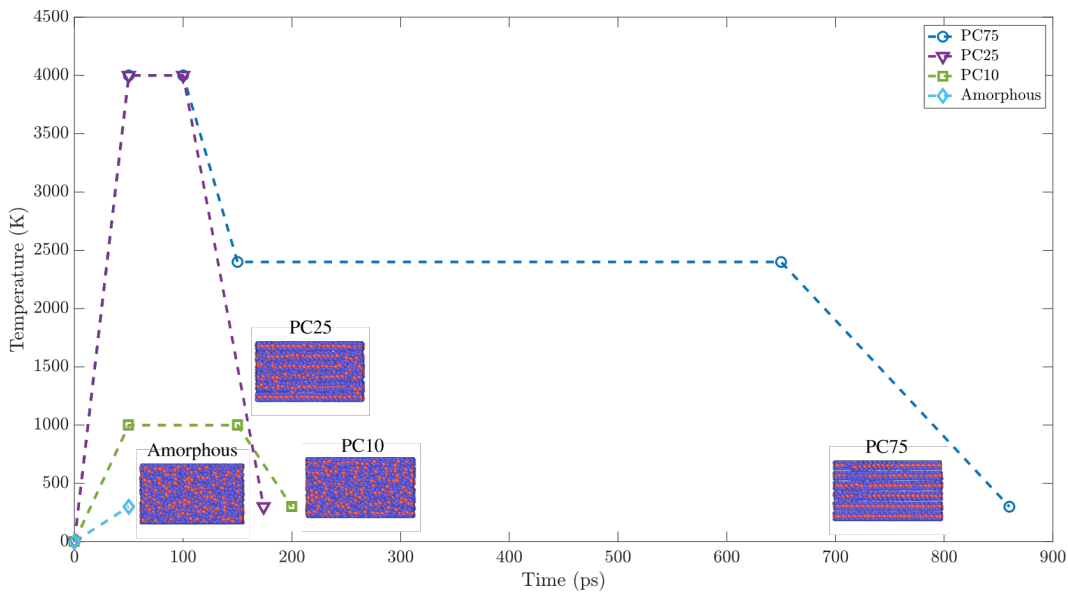

Figure 1: Sequence of annealing and quenching temperatures used to create polycrystalline models with different percentages of crystals

All MD simulations are performed using the LAMMPS open-source package [58], and the postprocessing of results are performed using OVITO [59] and MATLAB script. Atomic interactions were modeled using the reactive force field (ReaxFF) [60], implemented via the reax/c user package in LAMMPS [61]. The ReaxFF parameters employed in this study were specifically developed to accurately capture the interactions in Mo, S and carbon systems [62]. These parameters have been validated to reproduce the energetics of both crystalline Mo-S phases and amorphous $MoS_2$, making them well-suited for simulating the crystallization of $MoS_2$. While previous studies have utilized the reactive empirical bond order (REBO) potential to model atomic interactions [31], our prior work has demonstrated its limitations [63]. Specifically, the REBO potential fails to predict the enthalpies of formation for Mo-S crystalline phases other than $MoS_2$ and inaccurately estimates the relative energies of amorphous $MoS_2$ compared to its crystalline counterpart [46, 63]. These shortcomings underscore the superiority of ReaxFF parameters used for this study, justifying its selection over REBO.

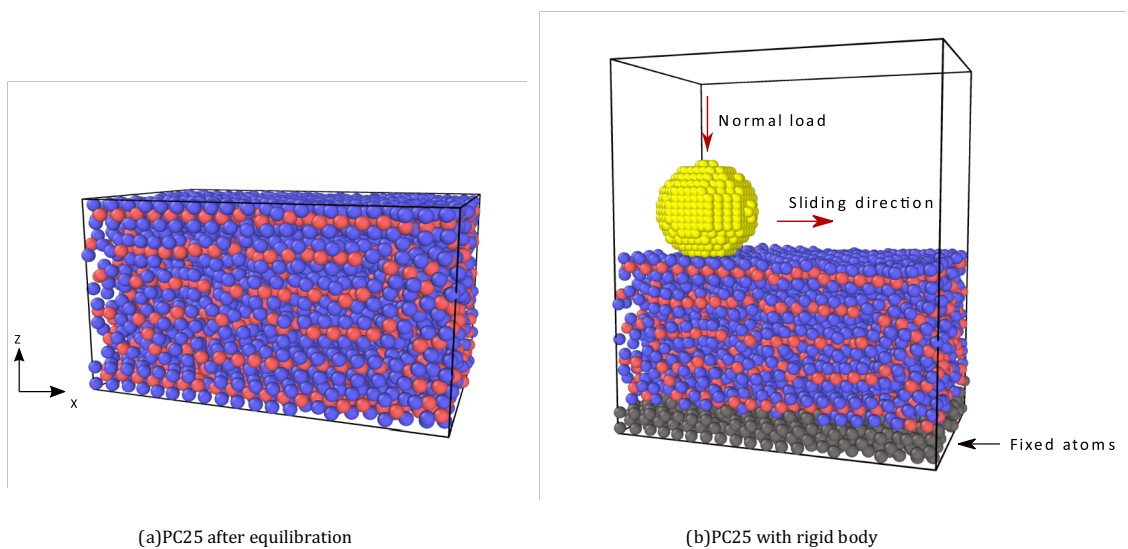

(a) PC25 after equilibration    (b) PC25 with rigid body

Figure 2: Atomic configuration of PC25, rigid body and imposed boundary conditions for sliding simulation

The simulation setup involves placing the rigid body on top of the $MoS_2$ surface, as illustrated in Figure 2b. Periodic boundary conditions are applied in the X and Y directions, while a nonperiodic boundary condition is used in the Z direction. A normal load is applied in the Z direction, and a constant horizontal sliding velocity of $10^{-4}$ Å /s is imposed in the X direction. Five different normal loads — 1.49, 2.98, 4.97, 5.46, and 9.94 nN — are considered to investigate the load-dependent frictional behavior. To ensure stability during sliding, the bottom portion of the $MoS_2$ model, consisting of approximately 900 atoms (grey colored balls in Figure 2b) are constrained in all directions. The movement of the rigid body is controlled using the *fix rigid* command in LAMMPS. The system is maintained at a constant temperature of 300 K using the Nosé - Hoover thermostat (NVT ensemble). Each sliding simulation is performed for 1 ns with a timestep of 0.5 fs. To study wear mechanisms, a higher normal load of 24.85 nN is applied while keeping the horizontal sliding velocity unchanged.

## 3. Results and discussion

In this section, we present and discuss the key findings from our study, focusing on the influence of crystallinity on the frictional and wear behavior of MoS$_2$. First, we compute the COF for different degrees of crystallinity, ranging from amorphous to crystalline structures. Second, we investigate the underlying friction mechanisms to provide atomic-scale insights. Finally, we examine wear mechanism observed under higher normal loads, highlighting the role of crystallinity in determining the wear resistance.

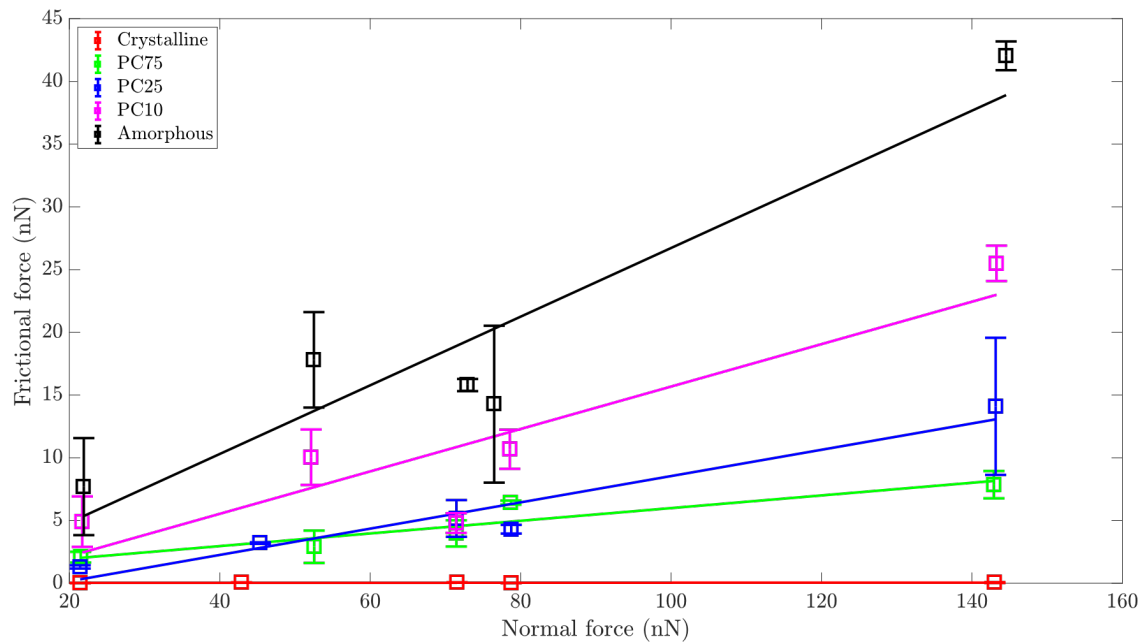

Figure 3: Plots of normal force with respect to frictional force for MoS$_2$ with different degrees of crystallinity to measure the coefficient of friction

### 3.1. Coefficient of friction

The COF is determined from the sliding simulations conducted at varying normal loads. For each applied normal load, the corresponding frictional forces are measured. To ensure the reliability of results, three independent MD runs are performed and the frictional forces are calculated as the average of these runs. The results are presented in Figure 3, where the frictional force is plotted with respect to the normal load. The error bars in the plots indicate the standard deviation across the three runs. The COF is then computed from the slope of the linear fit to the data, as illustrated in Figure 3. The calculated values of COF with corresponding error bounds are presented in Table 1.

The computed values of COF summarized in Table 1 reveals a clear dependence on the degree of crystallinity in MoS$_2$. The lowest and highest values are observed for crystalline and amorphous MoS$_2$ respectively. We obtained a COF value of 0.00014 for crystalline MoS$_2$, which falls in the category of superlubricity [18]. This result aligns with previous experimental and computational studies that have reported superlubricity [6, 18, 52]. However, while most prior studies focused on sliding $MoS_2$ over itself, our work demonstrates superlubricity in crystalline $MoS_2$ during rigid body sliding simulations. The COF value increases with a reduction in crystallinity. Notably, the highest value (0.2736) is observed for amorphous $MoS_2$, which is consistent with the values reported in [51]. This trend underscores the significant role of crystallinity in determining the frictional behavior of $MoS_2$, with higher crystallinity favouring lower friction. In the following section, we explore the nanoscale friction mechanisms responsible for this variation in COF with respect to the degree of crystallinity.

Table 1: Coefficient of friction for $MoS_2$ with different degrees of crystallinity

| System | Coefficient of friction |
|---|---|
| Crystalline | 0.00014±0.00005 |
| PC75 | 0.05±0.02 |
| PC25 | 0.11±0.02 |
| PC10 | 0.17±0.04 |
| Amorphous | 0.28±0.09 |

*3.2. Friction mechanism*

In this section, we explore the friction mechanisms responsible for the observed differences in COF with varying crystallinity. For crystalline $MoS_2$, the sliding surface remains flat throughout the simulation, even under the highest normal load of 9.94 nN. This flat surface minimizes resistance to the motion of rigid body, contributing to the exceptionally low COF value. Another major factor for the superlubricity observed in crystalline $MoS_2$ is the anisotropy in friction [64, 65]. Even though we slide the rigid body in horizontal direction, the lateral movement is not restricted, allowing the rigid body to follow a zig-zag path across the $MoS_2$ surface as shown in Figure 4. The movement of the rigid body is captured at four different frames, each representing a window of 0.125 ns. In Figure 4, the red arrow in the top two frames indicates the direction of sliding, while the black arrows depict the actual path followed by the rigid body. The movement of the rigid body over time is represented in a clockwise direction in Figure 4. This zig-zag trajectory represents the path of minimum energy, further reducing frictional resistance. The anisotropic friction observed in this work is consistent with the findings reported in prior literature [65]. However, this anistropic friction is less prominent in polycrystalline $MoS_2$ and almost negligible in amorphous $MoS_2$, thereby validating their high values of COF.

To further understand the variation in COF values, we investigate the surface roughness of $MoS_2$ as a function of crystallinity. Using the Gaussian density method in OVITO [59, 66], we construct a geometric surface mesh that describes the outer and inner boundaries of the atomic structure. This method generates an isosurface from a volumetric density field computed by superimposing 3D Gaussian functions centered at each atom in the top layer. The resulting mesh provides a quantitative estimate of surface roughness, which is visualized through color-coded gradients representing the distance from the bottom to the top of the surface. Here, we select atoms present at the top surface of $MoS_2$ and construct a surface mesh. The resulting surface meshing for MoS$_2$ with varying degrees of crystallinity are presented in Figures 5a-5d, illustrating the relationship between crystallinity and surface morphology. The red and blue colored regions in the figures represent the atoms at the highest and lowest positions among the group of selected atoms. A constant color in the plot implies that all the atoms are positioned at the same level. This is evident in Figures 5a & 5b (crystalline and PC25), where there is negligible color variation. In contrast, PC10 (Figure 5c) and amorphous MoS$_2$ (Figure 5d) exhibits pronounced color variations, denoting a significant surface disorder or roughness. This progressive surface roughness with decreasing crystallinity along with the anisotropy in friction explains the observed trend in COF values. Next, we investigate the effect of crystallinty on the wear resistance of MoS$_2$.

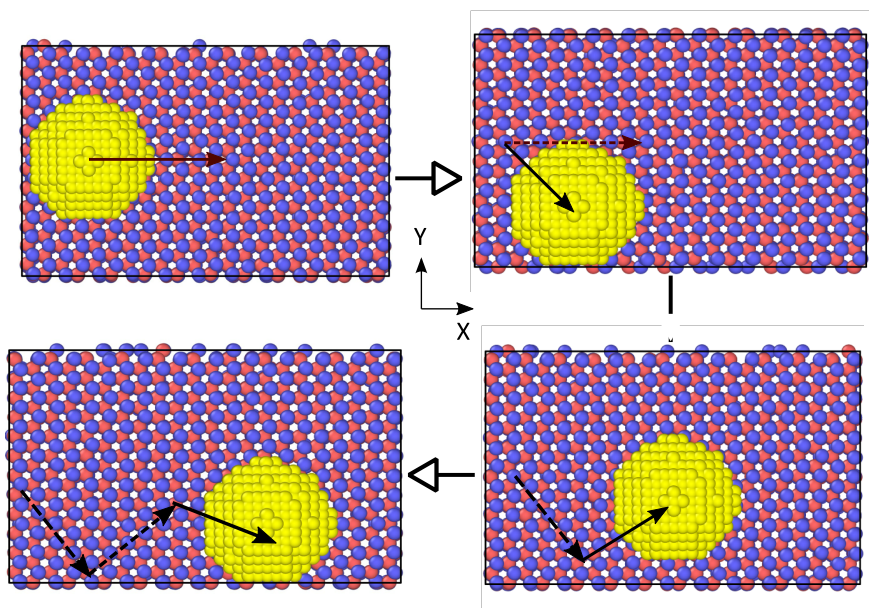

Figure 4: Zig-zag movement of rigid body during sliding simulation in crystalline MoS$_2$

*3.3. Wear characteristics*

To study the wear characteristics of MoS$_2$ with different levels of crystallinity, we performed sliding simulation at an elevated normal load of 24.85 nN while maintaining a constant horizontal velocity of $10^{-4}$ Å /s. The path followed by the rigid body and the configurations of the top layer of atoms are investigated in detail and presented in Figure 6. Further, we measured the depth of wear from the initial and final position of the rigid body. Figure 6 reveals a strong dependence between crystallinty and wear resistance of MoS$_2$. For crystalline MoS$_2$ (Figure 6a), a minimal wear depth of 5.32 Å with localized bond distortion at the topmost layer is observed. This wear depth increases progressively with decreasing crystallinity, culminating in complete structural disintegration for both for PC10 and amorphous MoS$_2$ (Figures 6d&6e). Notably, the amorphous configuration shows a wear depth of 14.8 Å which is approximately three times greater than that of crystalline MoS$_2$.

To further reinforce the dependence between crystallinty and wear resistance, we quantitatively analyzed the displacement of surface atoms in MoS$_2$. The displacement vectors in the Z direction are calculated from the atomic positions before and after sliding. The displacements are color coded and presented in Figure 7. The red and blue colors represent maximum and minimum displacements.

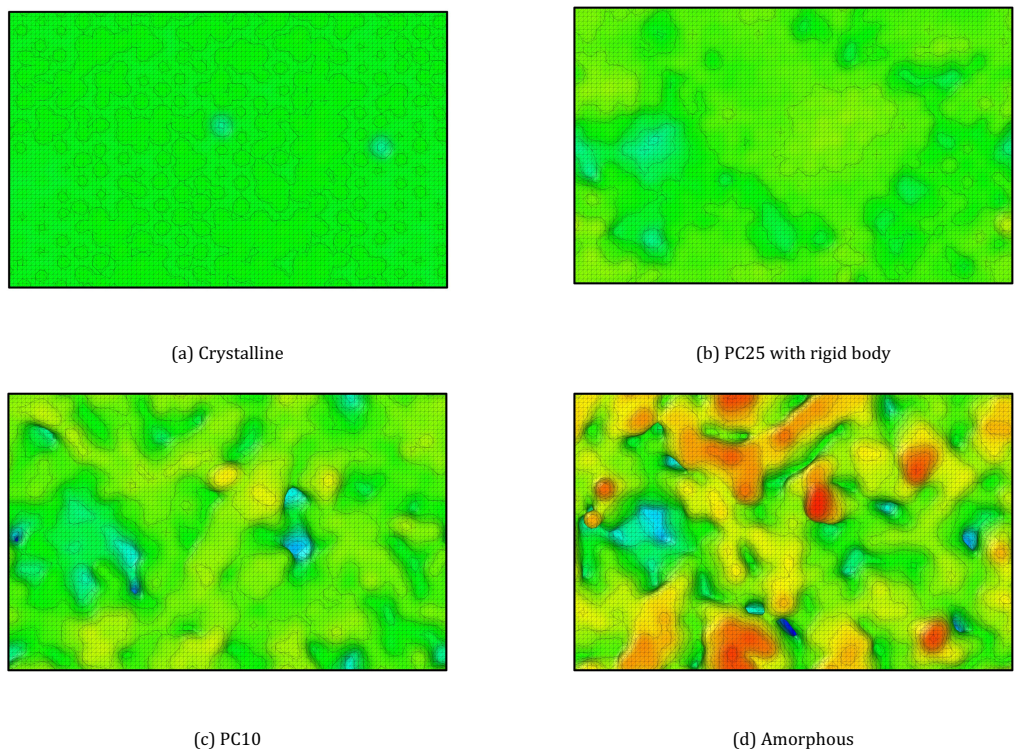

(a) Crystalline    (b) PC25 with rigid body

(c) PC10    (d) Amorphous

Figure 5: Geometric surface meshing for MoS$_2$ with different levels of crystallinity. Color gradients are used to represent the surface roughness with red and blue colored regions indicating atoms at the bottom at top surface.

Crystalline, PC75, and PC25 configurations (Figures 7a-7c) exhibit uniform atomic displacement patterns, indicating homogeneous deformation. In contrast, PC10 and amorphous MoS$_2$ (Figures 7d&7e), demonstrate significant displacement, with the amorphous structure showing distinct void formation at the surface. The pronounced displacement of atoms at lower degrees of crystallinity implies a reduction in wear resistance. These atomic-scale observations — combining both structural configurations and displacement patterns — establish a direct correlation between crystallinity and wear resistance. Thus, a higher degree of crystallinity provides superior wear resistance compared to amorphous MoS$_2$.

## 4. Conclusion

This study systematically investigates the effect of crystallinity on the frictional and wear performance of MoS$_2$ using rigid body sliding simulations. Our findings reveal that crystallinity plays a crucial role in determining the COF and wear resistance. Crystalline MoS$_2$ exhibits superlubricity, with an exceptionally low COF of 0.00014, due to its ability to maintain a flat sliding surface and leverage frictional anisotropy.

This allows for lateral movement along the minimum energy path, reducing resistance to sliding. In contrast, amorphous MoS$_2$ shows a significantly higher COF of 0.2736, attributed to increased surface roughness and atomic-scale hindrances.

Beyond friction, the study highlights the role of crystallinity in wear resistance. Under high normal loads, crystalline MoS$_2$ demonstrates superior resistance to material deformation and adhesive interactions, preserving its structural integrity. In contrast, amorphous MoS$_2$ is more susceptible to wear due to localized atomic rearrangements and increased energy dissipation at the contact interface. These results emphasize the importance of structural order in optimizing the tribological properties of MoS$_2$.

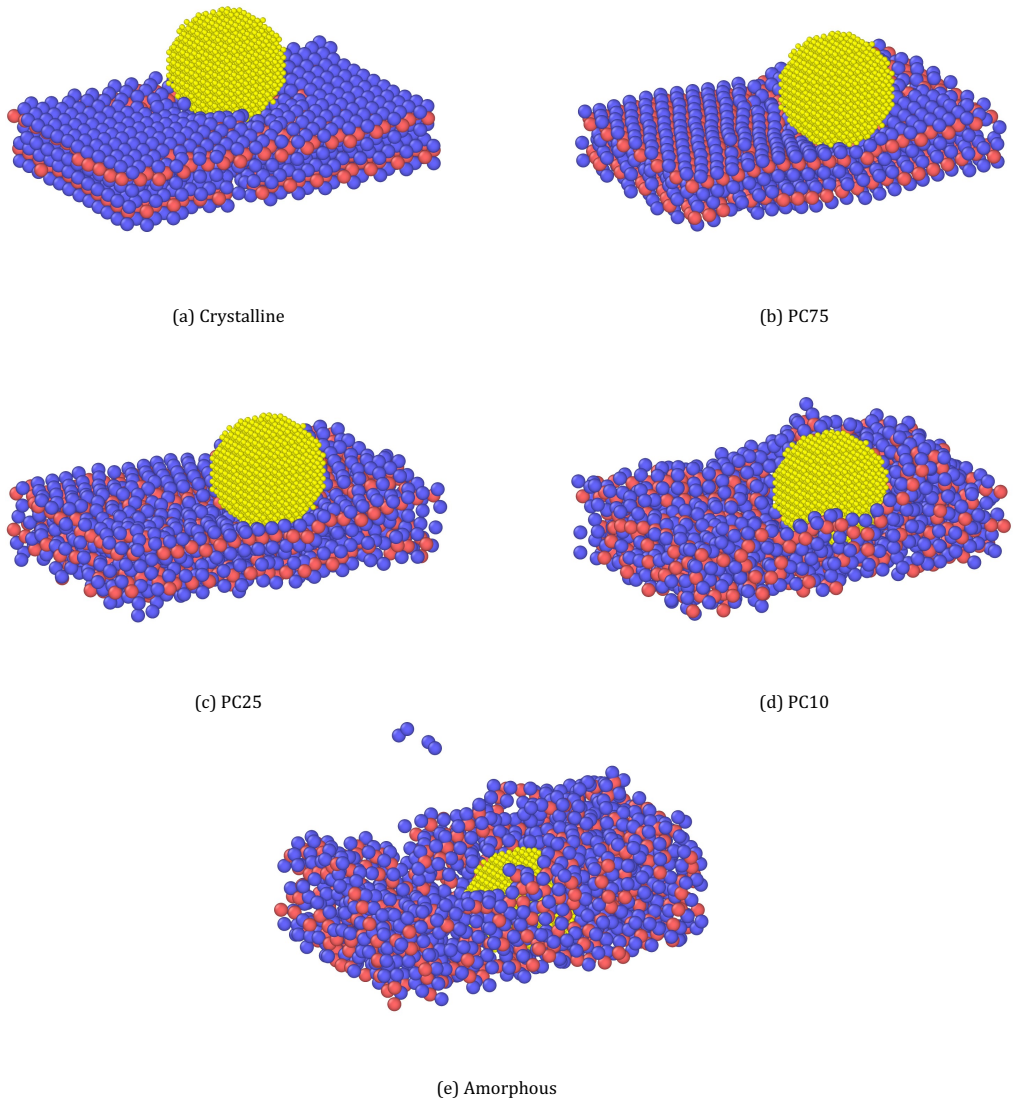

Figure 6: Atomic representation of wear in MoS$_2$ with different levels of crystallinity

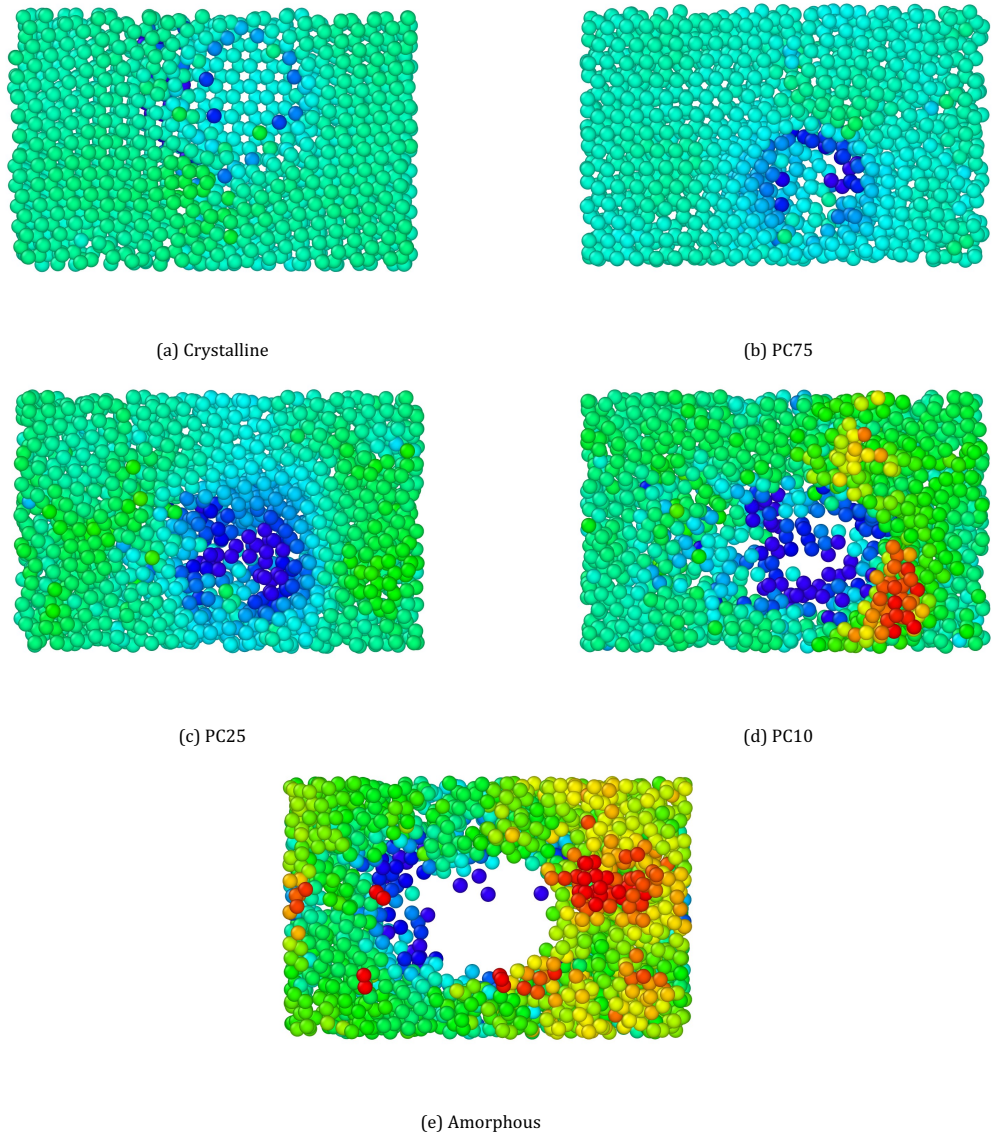

Figure 7: Color coded displacement of atoms at the top surface of MoS$_2$. Red and blue colors denote atoms with maximum and minimum displacement.

Our findings confirm previous reports of superlubricity in MoS$_2$ and extend the understanding to rigid-body sliding conditions. This work provides valuable atomic-scale insights into the friction and wear mechanisms of MoS$_2$, offering a fundamental basis for designing advanced solid lubricants and wear-resistant coatings. The results have direct implications for engineering applications requiring ultralow friction and high durability, such as aerospace, microelectromechanical systems, and next generation lubricating technologies.

**CRediT authorship contribution statement**

**Abhiram B.R:** Writing – review & editing, Writing – original draft, Conceptualization, Methodology, Investigation, Formal analysis. **Ilia Ponomarev:** Writing – review & editing, Conceptualization, Methodology, Investigation. **Tomas Polcar:** Writing – review & editing, Investigation, Supervision, Funding acquisition.

**Declaration of competing interest**

The authors declare that they have no known competing financial interests or personal relationships that could have appeared to influence the work reported in this paper.

**Acknowledgement**

The project was supported by the Czech Science Foundation (project No. 23-07785S); the computing part was partially co-funded by the European Union under the project Robotics and advanced industrial production (reg. no. CZ.02.01.01/00/22_008/0004590). This work was supported by the Ministry of Education, Youth and Sports of the Czech Republic through the e-INFRA CZ (ID:90254)

**Data availability**

Data will be made available on request.

**Appendix A. Degree of crystallinity**

The degree of crystallinity is quantified as the percentage of Mo atoms classified as "crystalline Mo" among all Mo atoms in the system. This classification is based on the characteristic structural arrangement of the 2H-$MoS_2$ crystal. To determine whether a given Mo atom is crystalline, we analyze its surrounding Mo neighbors and the angles formed by Mo-Mo-Mo triplets, where the atom in focus is the central one. In a perfect crystal, each Mo atom is surrounded by six second-neighbor Mo atoms arranged in the corners of a perfect hexagon. This results in six Mo-Mo-Mo angles of 60°, six angles of 120°, and three angles of 180°. To identify crystalline Mo atoms in our analysis, we consider atoms meeting the following criteria: at least five Mo-Mo-Mo angles must fall within 50° to 70°, at least five more within 110° to 130°, and at least two angles exceeding 170°. This criterion minimizes false positive identifications. Figure A.1 provides two examples of crystalline Mo atoms: one nearly ideal from a perfect crystal (left) and another highly distorted from an amorphous structure (right). It is to be noted that crystallinity can also be assessed using S atoms instead of Mo atoms. However, both methods provide equivalent degrees of crystallinity.

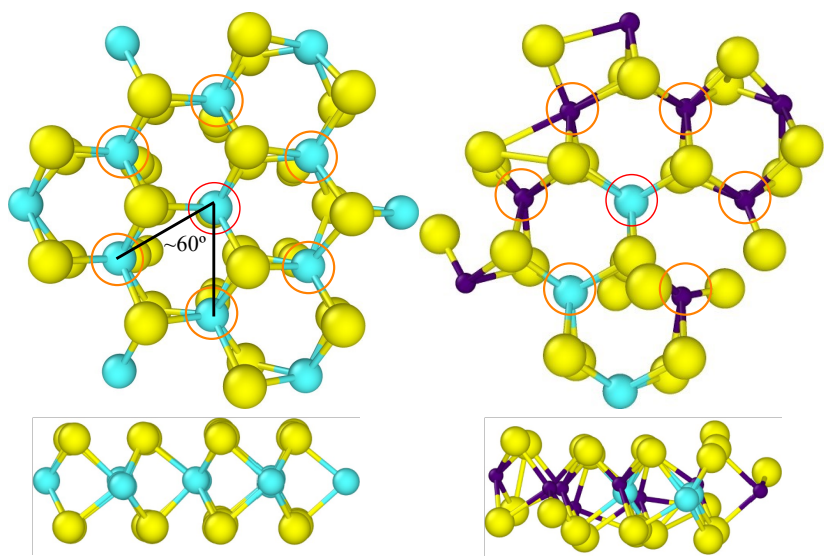

Figure A.1: Representation of "crystalline Mo" atoms: the top images show views along z-axis, while bottom images present views perpendicular to it. Cyan spheres are "crystalline Mo" atoms, purple spheres - other Mo atoms, yellow spheres - sulfur atoms. The red circle highlights the Mo atom in focus, orange circles denote second-neighbor Mo atoms.